\documentclass[a4paper]{article}

\usepackage{makecell}
\usepackage{xcolor}
\usepackage{url}
\usepackage{nicefrac}
\usepackage{booktabs}
\usepackage{multirow}
\usepackage{INTERSPEECH2019}
\usepackage{amssymb}
\usepackage{pifont}
\newcommand{\cmark}{\ding{51}}%
\newcommand{\xmark}{\ding{55}}%
\newcommand{\newpara}[1]{\vspace{6pt}\noindent\textbf{#1}}
\usepackage[implicit=false]{hyperref}

\title{Clova Baseline System for the VoxCeleb Speaker Recognition Challenge 2020}
\name{Hee Soo Heo, Bong-Jin Lee, Jaesung Huh, Joon Son Chung}
\address{Naver Corporation, South Korea 
\email{}}

\begin{document}

\maketitle
\begin{abstract}
This report describes our submission to the VoxCeleb Speaker Recognition Challenge (VoxSRC) at Interspeech 2020. We perform a careful analysis of speaker recognition models based on the popular ResNet architecture, and train a number of variants using a range of loss functions. Our results show significant improvements over most existing works without the use of model ensemble or post-processing. We release the training code and pre-trained models as unofficial baselines for this year's challenge.

\end{abstract}
\vspace{10pt}


\section{Introduction}

The VoxCeleb Speaker Recognition Challenge 2020 is second installment of the new series of speaker recognition challenges that are hosted annually. The challenge is intended to assess how well current speaker recognition technology is able to identify speakers in unconstrained or 'in the wild' data. This year's challenge is different to the last in a number of ways: (1) there is an explicit domain shift between the training data and the test data; (2) the test set contains utterances that are shorter than the segments seen during training.
The following sections of this report describe the method that underlies our submission to the challenge.

\section{Model}

\subsection{Input representation}

During training, we use a fixed length 2-second temporal segment extracted randomly from each utterance.
Pre-emphasis is applied to the input signal using a coefficient of 0.97.
Spectrograms are extracted with a hamming window of width 25ms and step 10ms with a FFT size of 512. 64-dimensional log Mel-filterbanks are used as the input to the network.
Mean and variance normalisation (MVN) is performed by applying instance normalisation~\cite{ulyanov2016instance} to the network input.
Since the VoxCeleb dataset consists mostly of continuous speech, voice activity detection is not used in training and testing.

\subsection{Trunk architecture}

Residual networks~\cite{he2016deep} are widely used in image recognition and have been applied to speaker recognition~\cite{Chung18a,cai2018exploring,Xie19a,chung2019delving}. We use two variants of the ResNet with 34 layers.

\newpara{Speed optimised model.} 
The first variant uses only one {\bf quarter} of the channels in each residual block compared to the original ResNet-34 in order to reduce computational cost. The model only has 1.4 million parameters compared to 22 million of the original ResNet-34. Self-attentive pooling (SAP)~\cite{cai2018exploring} is used to aggregate frame-level features into utterance-level representation while paying attention to the frames that are more informative for utterance-level speaker recognition. The network architecture is identical to that used in~\cite{chung2020defence} except for the input dimension, and we refer to this configuration as {\bf Q~/~SAP} in the results.

\newpara{Performance optimised model.} 
The second slower variant has {\bf half} of the channels in each residual block compared to the original ResNet-34, and contains 8.0 million parameters. Moreover, the stride at the first convolutional layer is removed, leading to increased computational requirement. Attentive Statistics Pooling (ASP)~\cite{okabe2018attentive} is used to aggregate temporal frames, where the channel-wise weighted standard deviation is calculated in addition to the weighted mean. 
Table~\ref{table:model} shows the detailed architecture of the performance optimised model.
We refer to this configuration as {\bf H~/~ASP} in the results. 


\begin{table}[h]
 \caption{Trunk architecture for the performance optimized model. {$\mathbf{L}$}: length of input sequence, {\bf ASP}: attentive statistics pooling.}
  \centering
  \label{table:model}
  \begin{tabular}{l c c c}
  
  \toprule

   Layer & Kernel size  & Stride& Output shape   \\
  \hline
  Conv1 & $3 \times 3 \times 32$ & $1\times1$& $L \times 64\times 32$ \\
  \hline
  Res1 & $3\times 3 \times 32$ & $1\times1$ & $ L \times64 \times 32$ \\
  \hline
  Res2 & $3\times 3\times 64$ &$2\times2$ & $ \nicefrac{L}{2} \times32 \times 64$ \\
  \hline
  Res3 & $3\times 3 \times 128$ &$2\times2$ & $ \nicefrac{L}{4} \times16 \times 128$\\
  \hline
  Res4 & $3\times 3\times 256$ &$2\times2$ & $ \nicefrac{L}{8} \times8 \times 256$\\
  \hline
  Flatten & - & - & $ \nicefrac{L}{8} \times 2048$\\
  \hline
  ASP & - & - & $ 4096$ \\
  \hline
  Linear & $ 512 $ & - & $ 512 $ \\
  \bottomrule
  \end{tabular}
\end{table}
\begin{table*}[h]
\caption{Results on the VoxCeleb and VoxSRC test sets.
The figures in bold represent the best results for each metric, excluding the fusion outputs.
{\bf AP:} Angular Prototypical.
{\bf BN:} Batch Normalisation on the speaker embeddings.
$\dag$ This method uses score normalisation as a post-processing step.}
\label{table:results}
\footnotesize
\setlength{\tabcolsep}{4pt}
\renewcommand\arraystretch{1.2}
\centering
\begin{tabular}{ l l l l | r  r | r r | r r | r r | r r  }
\toprule
 \textbf{Config.} & \textbf{Loss} & \textbf{Aug.} & \textbf{BN}   & \multicolumn{2} {c|} {\bf VoxCeleb1}& \multicolumn{2} {c|} {\bf VoxCeleb1-E cl.} & \multicolumn{2} {c|} {\bf VoxCeleb1-H cl.} & \multicolumn{2} {c|} {\bf VoxSRC 2019} & \multicolumn{2} {c} {\bf VoxSRC 2020 Val} \\
    &     &   & & \textbf{EER} & \textbf{MinDCF} & \textbf{EER} & \textbf{MinDCF}  & \textbf{EER} & \textbf{MinDCF}  & \textbf{EER} & \textbf{MinDCF} & \textbf{EER} & \textbf{MinDCF} \\ 
\midrule

FR-34~\cite{chung2020defence} & AP           & \xmark & \xmark & 2.22 & - & - & - & - & - & - & -           & - & - \\
Sys 1~\cite{zeinali2019but} & Softmax~$\dag$ & \cmark & \xmark & - & - & 1.35 & - & 2.48 & - & - & -        & - & - \\
Fusion~\cite{zeinali2019but} & - & - & - & - & - & 1.14 & - & 2.21 & - & 1.42 & -        & - & - \\
Sys A5~\cite{garcia2020jhu} & AM-Softmax & \cmark & \xmark & - & - & 1.51 & - & - & - & 1.72 & -        & - & - \\
Fusion~\cite{garcia2020jhu} & - & - & - & - & - & 1.22 & - & - & - & 1.54 & -        & - & - 
\\ \midrule

Q / SAP & AM-Softmax & \xmark & \cmark &  2.20  & 0.139 &          2.10 & 0.137 &     3.67 & 0.213  &    2.31 & 0.144         & 5.43 & 0.295  \\
Q / SAP & AAM-Softmax & \xmark & \cmark &  2.13 & 0.138 &          2.12 & 0.140 &     3.52 & 0.211  &    2.33 & 0.145         & 5.19 & 0.290  \\
Q / SAP & AP          & \xmark & \cmark &  1.90 & 0.133 &          1.99 & 0.144 &     3.80 & 0.243  &    2.24 & 0.152         & 5.67 & 0.325 \\
Q / SAP & AP+Softmax & \xmark & \cmark &  1.85  & 0.119 &          1.96 & 0.138 &     3.65 & 0.233  &    2.16 & 0.146         & 5.49 & 0.311  \\ \midrule

H / ASP & AM-Softmax & \cmark & \xmark &  1.64  & 0.115 &       1.67 & 0.114 &     3.07 & 0.191  &     1.88 & 0.112          & 4.59 & 0.263 \\
H / ASP & AAM-Softmax & \cmark & \xmark &  1.59  & 0.113 &       1.50 & 0.105 &    2.91 & 0.181  &    1.74 & 0.103          & 4.40 & 0.244  \\
H / ASP & AP          & \cmark & \xmark & 1.50  & 0.126 &       1.69 & 0.120 &     3.39 & 0.217  &    1.92 & 0.128          & 5.09 & 0.288 \\
H / ASP & AP+Softmax  & \cmark & \xmark &  {\bf 1.18}  & {\bf 0.086} &      {\bf 1.21} & {\bf 0.086} &     {\bf 2.38} & {\bf 0.154}  &     {\bf 1.46} & {\bf 0.088}         & {\bf 3.79} & {\bf 0.213} \\ \midrule

H / ASP & AM-Softmax & \cmark & \cmark &  1.49  & 0.104 &       1.45 & 0.099 &     2.64 & 0.165  &     1.64 & 0.097         & 4.05 & 0.231  \\
H / ASP & AAM-Softmax & \cmark & \cmark &  1.28  & {\bf 0.086} &      1.34 & 0.091 &     2.48 & 0.155  &    1.61 & 0.093          & 3.81 & 0.218 \\
H / ASP & AP          & \cmark & \cmark & 1.43  & 0.113 &       1.45 & 0.105 &     2.91 & 0.189  &    1.74 & 0.117          & 4.43 & 0.262 \\
H / ASP & AP+Softmax  & \cmark & \cmark &  1.25  & 0.087 &      1.34 & 0.095 &     2.71 & 0.175  &     1.49 & 0.102         & 4.24 & 0.243  \\
 \bottomrule
\end{tabular} 
\end{table*}

\subsection{Loss function}

We train the networks using various types of loss functions widely used in speaker recognition. 
Additive margin softmax (AM-softmax)~\cite{wang2018cosface} and Additive angular margin softmax (AAM-softmax)~\cite{deng2019arcface} loss functions have been proposed in face recognition and successfully applied to speaker recognition~\cite{xiang2019margin}. These functions introduce a concept of margin between classes where the margin increases inter-class variance.
For both AM-Softmax and AAM-Softmax loss functions, we use a margin of 0.2 and a scale of 30 since this value results in the best performance on the VoxCeleb1 test set.

Angular Prototypical (AP) loss, a variant of the prototypical networks with an angular objective, has been used in~\cite{chung2020defence}, where it has demonstrated strong performance without manually-defined hyper-parameters.

\begin{figure}[!htb]
\centering 
\includegraphics[width=0.7\columnwidth]{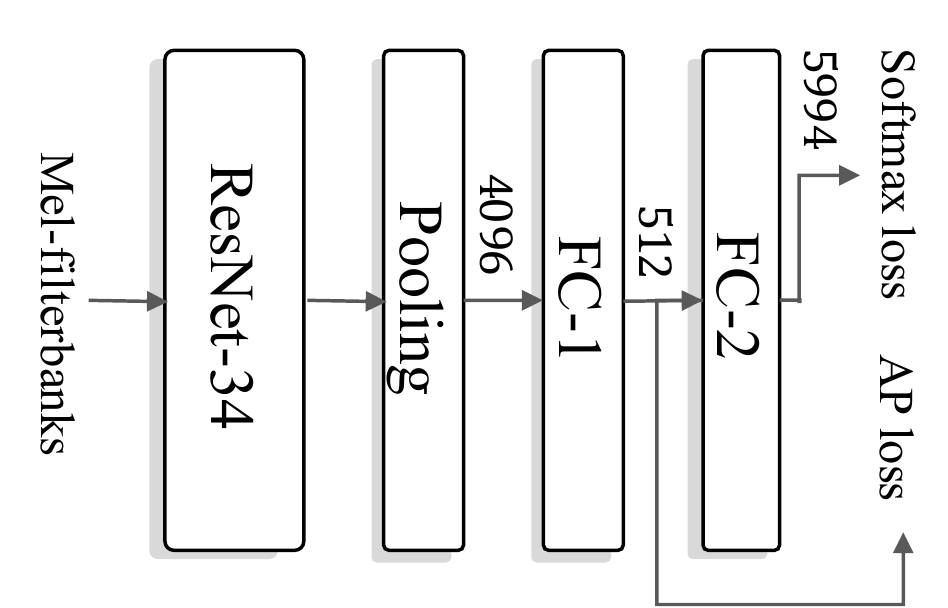}
\caption{Overview of the model architecture and the training strategy of AP+Softmax systems.}
\label{fig:overview} 
\end{figure}

Finally, we combine the Angular Prototypical loss with the vanilla softmax loss which demonstrates further improvement over using each of the loss functions. Figure~\ref{fig:overview} shows the model architecture and the training strategy for combining the AP and softmax loss functions.

\section{Experiments}

\subsection{Dataset}

The models are trained on the development set of VoxCeleb2~\cite{Chung18a}, which contains 5,994 speakers. The VoxCeleb1 test sets~\cite{Nagrani17} and the previous year's VoxSRC test set~\cite{chung2019voxsrc} are used as validation sets.

\subsection{Data augmentation}

We exploit two popular augmentation methods in speech processing -- additive noise and room impulse response (RIR) simulation. For additive noise, we use the MUSAN corpus~\cite{snyder2015musan} which contains 60 hours of human speech, 42 hours of music, and 6 hours of other noises such as dialtones or ambient sounds. For room impulse responses, we use the simulated RIR filters provided in~\cite{ko2017study}. Both noise and RIR filters are randomly selected in every training step.
 
Types of augmentation used are similar to ~\cite{snyder2018x}, in which the recordings are augmented by one of the following methods.
 
 \begin{itemize}
 \item \textbf{Speech}: Three to seven recordings are randomly picked from MUSAN speech, then added to the original signal with random signal to noise ratio (SNR) from 13 to 20dB. The duration of additive noise is matched to the sampled segment.
 \item \textbf{Music}: A single music file is randomly selected from MUSAN, and added to the original signal with a similar fashion from 5 to 15dB SNR.
 \item \textbf{Noise}: Background noises without human speech and music in MUSAN are added to the recording from 0 to 15dB SNR.
 \item \textbf{RIR filters}: Speech reverberation is performed via convolution operation with a collection of simulated RIR filters. We vary the gain of RIR filters to make more diverse reverberated signals.
 
\end{itemize}

\subsection{Implementation details}

Our implementation is based on the PyTorch framework~\cite{paszke2019pytorch} and trained on the NAVER Smart Machine Learning (NSML) platform~\cite{sung2017nsml}.
The models are trained using 8 NVIDIA P40 GPUs with 24GB memory with the Adam optimiser. We use the distributed training implementation of \url{https://github.com/clovaai/voxceleb_trainer} where one epoch is defined as a full pass through the dataset {\em by each GPU}.

\newpara{Speed optimised model.} 
We use an initial learning rate of 0.01, reduced by $10\%$ every 2 epochs. The network is trained for 50 epochs. 
We use a mini-batch size of 500.
The models take around 1 day to train.

\newpara{Performance optimised model.} 
We use an initial learning rate of 0.001, reduced by $25\%$ every 3 epochs. The network is trained for 36 epochs. A weight decay of {\em 5e-5} is applied.
We use a smaller batch size of 150 due to memory limitations.
The models take around 2 days to train.

\subsection{Scoring}

The trained networks are evaluated on the VoxCeleb1 and the VoxSRC test sets. We sample ten 4-second temporal crops at regular intervals from each test segment, and compute the $10 \times 10 = 100$ cosine similarities between the possible combinations
 from every pair of segments. The mean of the 100 similarities is used as the score. This protocol is in line with that used by~\cite{Chung18a,chung2019delving,chung2020defence}.

\subsection{Evaluation protocol}

We report two performance metrics: (i) the Equal Error Rate (EER) which is the rate at which both acceptance and rejection errors are equal; and (ii) the minimum detection cost of the function used by the NIST SRE~\cite{nist2018}
and the VoxSRC\footnote{\url{http://www.robots.ox.ac.uk/~vgg/data/voxceleb/competition2020.html}} evaluations. 
The parameters $C_{miss}=1$, $C_{fa}=1$ and $P_{target}=0.05$ are used for the cost function.

\subsection{Results}

Table~\ref{table:results} reports the experimental results. 

We compare our models to the two best scoring submissions~\cite{zeinali2019but,garcia2020jhu} in the VoxSRC 2019. From each of these submissions, we report the results of the best single model and the best fusion output.

The results demonstrate that the sum of metric learning and classification-based losses work best in most scenarios. The batch normalisation layer applied to the output contributes a significant improvement in performance for the classification objectives.

The performance optimised model trained with the {\bf AP+Softmax} loss and without the embedding batch normalisation produces an EER of 5.19\% and MinDCF of 0.314 on the VoxSRC 2020 test set.

\section{Conclusion}

The report describes our baseline system for the 2020 VoxSRC Speaker Recognition Challenge. The proposed system is trained using a combination of metric learning and classification-based objectives. Our best model outperforms all single model systems and all but one ensemble system in the last year's challenge. We release the full training code and pre-trained models as unofficial baselines for the challenge.


\newpage
\raggedbottom
\bibliographystyle{IEEEtran}
\bibliography{shortstrings,mybib}

\begin{thebibliography}{10}
\providecommand{\url}[1]{#1}
\csname url@samestyle\endcsname
\providecommand{\newblock}{\relax}
\providecommand{\bibinfo}[2]{#2}
\providecommand{\BIBentrySTDinterwordspacing}{\spaceskip=0pt\relax}
\providecommand{\BIBentryALTinterwordstretchfactor}{4}
\providecommand{\BIBentryALTinterwordspacing}{\spaceskip=\fontdimen2\font plus
\BIBentryALTinterwordstretchfactor\fontdimen3\font minus
  \fontdimen4\font\relax}
\providecommand{\BIBforeignlanguage}[2]{{%
\expandafter\ifx\csname l@#1\endcsname\relax
\typeout{** WARNING: IEEEtran.bst: No hyphenation pattern has been}%
\typeout{** loaded for the language `#1'. Using the pattern for}%
\typeout{** the default language instead.}%
\else
\language=\csname l@#1\endcsname
\fi
#2}}
\providecommand{\BIBdecl}{\relax}
\BIBdecl

\bibitem{ulyanov2016instance}
D.~Ulyanov, A.~Vedaldi, and V.~Lempitsky, ``Instance normalization: The missing
  ingredient for fast stylization,'' \emph{arXiv preprint arXiv:1607.08022},
  2016.

\bibitem{he2016deep}
K.~He, X.~Zhang, S.~Ren, and J.~Sun, ``Deep residual learning for image
  recognition,'' in \emph{Proc. CVPR}, 2016, pp. 770--778.

\bibitem{Chung18a}
J.~S. Chung, A.~Nagrani, and A.~Zisserman, ``{VoxCeleb2}: Deep speaker
  recognition,'' in \emph{Proc. Interspeech}, 2018.

\bibitem{cai2018exploring}
W.~Cai, J.~Chen, and M.~Li, ``Exploring the encoding layer and loss function in
  end-to-end speaker and language recognition system,'' in \emph{Speaker
  Odyssey}, 2018.

\bibitem{Xie19a}
W.~Xie, A.~Nagrani, J.~S. Chung, and A.~Zisserman, ``Utterance-level
  aggregation for speaker recognition in the wild,'' in \emph{Proc. ICASSP},
  2019.

\bibitem{chung2019delving}
J.~S. Chung, J.~Huh, and S.~Mun, ``Delving into {VoxCeleb}: environment
  invariant speaker recognition,'' in \emph{Speaker Odyssey}, 2020.

\bibitem{chung2020defence}
J.~S. Chung, J.~Huh, S.~Mun, M.~Lee, H.~S. Heo, S.~Choe, C.~Ham, S.~Jung, B.-J.
  Lee, and I.~Han, ``In defence of metric learning for speaker recognition,''
  in \emph{Proc. Interspeech}, 2020.

\bibitem{okabe2018attentive}
K.~Okabe, T.~Koshinaka, and K.~Shinoda, ``Attentive statistics pooling for deep
  speaker embedding,'' in \emph{Proc. Interspeech}, 2018.

\bibitem{zeinali2019but}
H.~Zeinali, S.~Wang, A.~Silnova, P.~Mat{\v{e}}jka, and O.~Plchot, ``{BUT}
  system description to voxceleb speaker recognition challenge 2019,''
  \emph{arXiv preprint arXiv:1910.12592}, 2019.

\bibitem{garcia2020jhu}
D.~Garcia-Romero, A.~McCree, D.~Snyder, and G.~Sell, ``{JHU-HLTCOE} system for
  the {VoxSRC} speaker recognition challenge,'' in \emph{Proc. ICASSP}.\hskip
  1em plus 0.5em minus 0.4em\relax IEEE, 2020, pp. 7559--7563.

\bibitem{wang2018cosface}
H.~Wang, Y.~Wang, Z.~Zhou, X.~Ji, D.~Gong, J.~Zhou, Z.~Li, and W.~Liu,
  ``Cosface: Large margin cosine loss for deep face recognition,'' in
  \emph{Proc. CVPR}, 2018, pp. 5265--5274.

\bibitem{deng2019arcface}
J.~Deng, J.~Guo, N.~Xue, and S.~Zafeiriou, ``Arcface: Additive angular margin
  loss for deep face recognition,'' in \emph{Proc. CVPR}, 2019, pp. 4690--4699.

\bibitem{xiang2019margin}
X.~Xiang, S.~Wang, H.~Huang, Y.~Qian, and K.~Yu, ``Margin matters: Towards more
  discriminative deep neural network embeddings for speaker recognition,'' in
  \emph{Asia-Pacific Signal and Information Processing Association Annual
  Summit and Conference}.\hskip 1em plus 0.5em minus 0.4em\relax IEEE, 2019,
  pp. 1652--1656.

\bibitem{Nagrani17}
A.~Nagrani, J.~S. Chung, and A.~Zisserman, ``{VoxCeleb}: a large-scale speaker
  identification dataset,'' in \emph{Proc. Interspeech}, 2017.

\bibitem{chung2019voxsrc}
J.~S. Chung, A.~Nagrani, E.~Coto, W.~Xie, M.~McLaren, D.~A. Reynolds, and
  A.~Zisserman, ``{VoxSRC} 2019: The first {VoxCeleb} speaker recognition
  challenge,'' \emph{arXiv preprint arXiv:1912.02522}, 2019.

\bibitem{snyder2015musan}
D.~Snyder, G.~Chen, and D.~Povey, ``Musan: A music, speech, and noise corpus,''
  \emph{arXiv preprint arXiv:1510.08484}, 2015.

\bibitem{ko2017study}
T.~Ko, V.~Peddinti, D.~Povey, M.~L. Seltzer, and S.~Khudanpur, ``A study on
  data augmentation of reverberant speech for robust speech recognition,'' in
  \emph{IEEE International Conference on Acoustics, Speech and Signal
  Processing}, 2017, pp. 5220--5224.

\bibitem{snyder2018x}
D.~Snyder, D.~Garcia-Romero, G.~Sell, D.~Povey, and S.~Khudanpur, ``X-vectors:
  Robust dnn embeddings for speaker recognition,'' in \emph{Proc.
  ICASSP}.\hskip 1em plus 0.5em minus 0.4em\relax IEEE, 2018, pp. 5329--5333.

\bibitem{paszke2019pytorch}
A.~Paszke, S.~Gross, F.~Massa, A.~Lerer, J.~Bradbury, G.~Chanan, T.~Killeen,
  Z.~Lin, N.~Gimelshein, L.~Antiga \emph{et~al.}, ``Pytorch: An imperative
  style, high-performance deep learning library,'' in \emph{NIPS}, 2019, pp.
  8024--8035.

\bibitem{sung2017nsml}
N.~Sung, M.~Kim, H.~Jo, Y.~Yang, J.~Kim, L.~Lausen, Y.~Kim, G.~Lee, D.~Kwak,
  J.-W. Ha \emph{et~al.}, ``Nsml: A machine learning platform that enables you
  to focus on your models,'' \emph{arXiv preprint arXiv:1712.05902}, 2017.

\bibitem{nist2018}
\emph{NIST 2018 Speaker Recognition Evaluation Plan}, 2018 (accessed 31 July
  2020),
  \url{https://www.nist.gov/system/files/documents/2018/08/17/sre18_eval_plan_2018-05-31_v6.pdf},
  See Section 3.1.

\end{thebibliography}
\end{document}